\documentclass{article}
\usepackage{spconf,amsmath,graphicx}
\usepackage[font=small,labelfont=bf]{caption}
\usepackage{algorithm}
\usepackage{algpseudocode}
\usepackage{lipsum}
\usepackage{indentfirst}
\usepackage{float}
\usepackage{graphicx}
\usepackage{subcaption}
\usepackage{adjustbox}
\usepackage{siunitx}
\usepackage{color}
\usepackage{amsfonts}
\usepackage{booktabs}
\title{$\mathcal{G}$\MakeLowercase{lidar}3DJ: A View-Invariant gait identification via flash lidar data correction}
\name{
\parbox{\linewidth}{\centering
Nasrin Sadeghzadehyazdi$^{\dagger}$,
~Tamal Batabyal$^{\dagger}$,
A.~Glandon$^{\dagger\dagger}$,
\textit{Nibir K. Dhar}$^{\ddagger}$, \textit{B.~O.~Familoni}$^{\ddagger}$, \textit{K.~M.~Iftekharuddin}$^{\dagger\dagger}$, \textit{Scott T. Acton}$^{\dagger}$ }
}

\address{$^{\dagger}$Department of Electrical and Computer Engineering, University of Virginia,\\ Charlottesville, USA, $^{\ddagger}$Night Vision and Electronic Sensors Directorate 
Fort Belvoir, USA,\\ $^{\dagger\dagger}$Electrical and Computer Engineering, Old Dominion University,
Norfolk, USA}

\begin{document}
%\ninept
%
\maketitle
\begin{abstract}
Gait recognition is a leading remote-based identification method, suitable for real-world surveillance and medical applications. Model-based gait recognition methods have been particularly recognized due to their scale and view-invariant properties. We present the first model-based gait recognition methodology, $\mathcal{G}$lidar3DJ using a skeleton model extracted from sequences generated by a single flash lidar camera. Existing successful model-based approaches take advantage of high quality skeleton data collected by Kinect and Mocap, for example, are not practicable for application outside the laboratory. The low resolution and noisy imaging process of lidar negatively affects the performance of state-of-the-art skeleton-based systems, generating a significant number of outlier skeletons. We propose a rule-based filtering mechanism that adopts robust statistics to correct for skeleton joint measurements. Quantitative measurements validate the efficacy of the proposed method in improving gait recognition.

\end{abstract}
{\let\thefootnote\relax\footnote{{This paper is accepted to be published in: 2019 IEEE International Conference on Image Processing, Sept 22-25, 2019, Taipei, Taiwan. 

IEEE  Copyright  Notice: {\scriptsize\textcircled{c}}IEEE  2019  Personal  use  of  this  material  is permitted. Permission from IEEE must be obtained for all other uses, in any current or future media,  including reprinting/republishing this material for advertising or promotional purposes, creating new collective works, for resale or redistribution to servers or lists, or reuse of any copyrighted component of this work in other works.}}}

\vspace{-0.3cm}
\section{Introduction}
\label{sec:intro}
\vspace{-0.3cm}

Gait recognition has been an active area of research in the last decade due to the widespread application in forensic cases, surveillance, and medical studies of patients affected by motion-related diseases like Parkinson's disease~\cite{del2016validation}. Gait recognition uses the features from both structure and motion for person identification. Previous studies have shown the relative uniqueness of gait for individuals \cite{cutting1977recognizing}. Unlike other biometric features such as the iris, face, and fingerprint, gait recognition does not require a subject's cooperation, nor does it require high quality data. Under uncontrolled real-world conditions, there are many scenarios in which direct contact between subjects and sensors is not possible, or there is a considerable distance between cameras and subjects that makes reliable data acquisition difficult or impossible. Under such conditions, many biometric methods fail; whereas, several studies have shown promising results for person identification with the gait-based biometric features \cite{lee2002gait,sinha2013person}. Furthermore, unlike color and texture, which are among the prevalent features in many identification studies, features extracted from gait are resilient to changes in clothing and lighting conditions.
\begin{figure}[!t]
    \centering
    \includegraphics[width=\linewidth,height=2.5in]{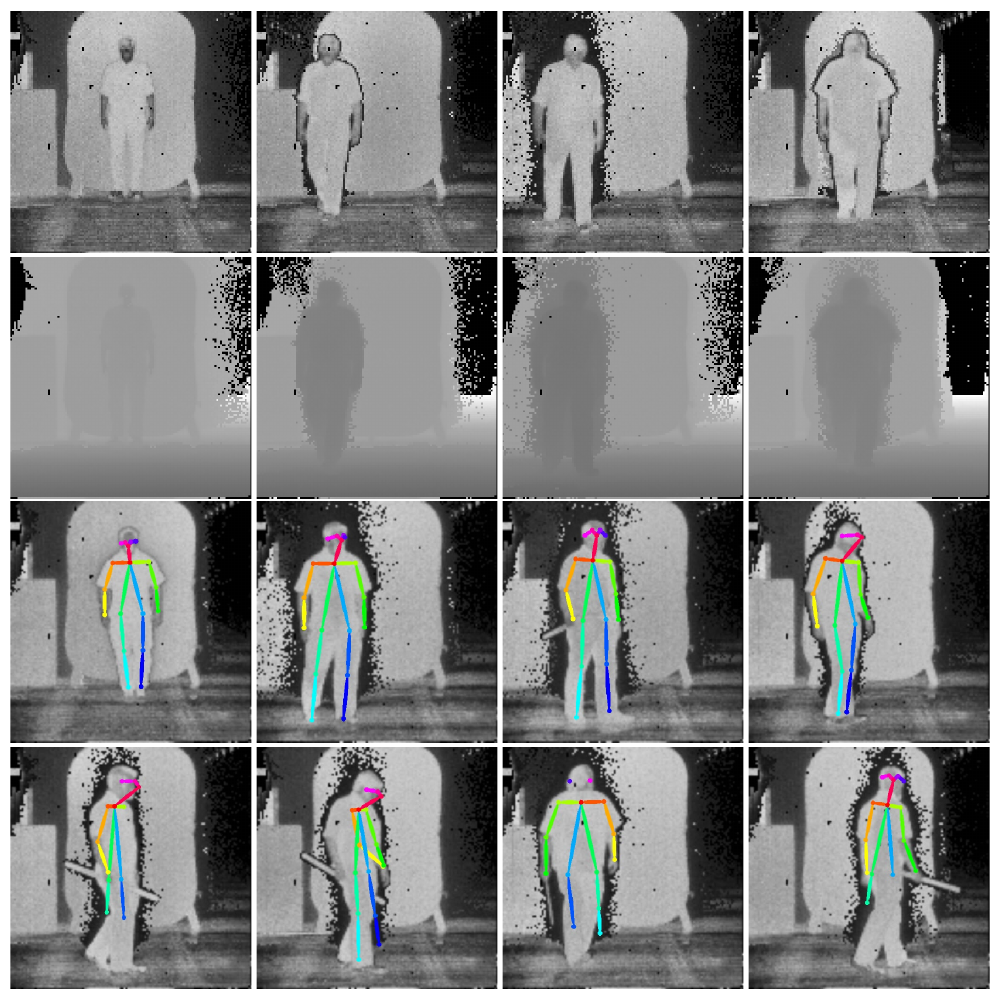}
    \makeatletter
    \vspace{-0.6cm}
    \caption[Sample frames of lidar]{Sample frames of lidar data. From top, first row: intensity data, second row: range data, third row: sample frames with correctly detected skeleton, last row: frames with faulty skeletons}    
    \label{fig:sampleFramesOfLidar}
    \vspace{-0.7cm}
\end{figure}

In recent years, depth-sensing cameras such as Kinect and lidar have become popular for gait analysis due to their ability to provide range (depth) and intensity data \cite{batabyal2016ugrad,batabyal2015ugrasp,clark2013concurrent,ofli2014sequence,batabyal2015action}. Each pixel in the range data provides distance information, so three-dimensional information (with the additional range dimension) can be recorded in time. Unlike ordinary cameras, the performance of depth cameras is not affected by changes in lighting conditions. In this work, we use data that was collected by a single flash lidar camera. A lidar sensor is a time-of-flight camera that uses laser beam to measure the distance of targets from the camera. A laser beam can be focused into small spots to suit the objects of interest and does not expand considerably by the object's surface, which gives a lidar sensor the capability of providing detailed images of a scene. As a result of such properties, lidar sensors have found application in areas such as archaeology, forestry, geology, geography, space missions, transportation and autonomous vehicles.
\begin{figure}[!t]
\centering
\resizebox{\linewidth}{!}{
\includegraphics[width=.95\textwidth]{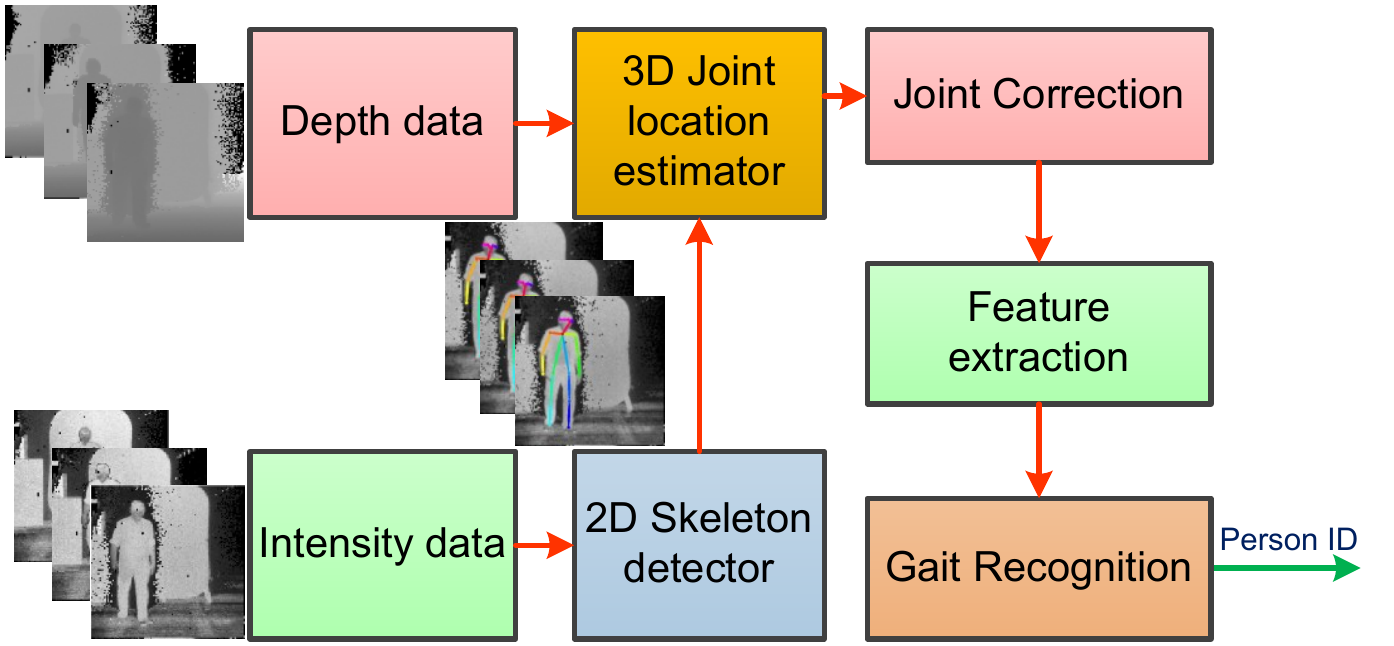}}
\vspace{-0.6cm}
\caption[work flowchart] {Pipeline for person identification using joint location correction}   
\label{fig:flowChart}
\vspace{-0.65cm}
\end{figure}

\vspace{-.35cm}
Gait recognition methods using video data are generally divided into two main categories, model-free methods and model-based methods. Model-free approaches require features from clean silhouettes \cite{han2006individual, rida2016human}. Model-based methods fit a model, like a skeleton, to human body and exploit features from the fitted model for recognition. Unlike model-free methods, model-based approaches are view and scale-invariant suited for real-world scenarios. On the other hand, model-based methods are computationally expensive. But, with depth-sensing modalities like Kinect that provide a direct estimation of joint positions, this expense is not an issue. However, Kinect sensors have the issues of limited range and unreliability of range information in outdoor environments, in particular under direct sunlight, where the high intensity infrared of environment cannot be easily differentiated from the infrared light of the sensor \cite{fankhauser2015kinect}. Compared with Kinect, a flash lidar camera has a drastically extended range ($>1000$~meters) and its performance is not affected in outdoor environments due to the high irradiance power of pulsed laser compared with the background \cite{horaud2016overview}.
\vspace{-.3cm}
\begin{figure}[!ht]
\centering
\includegraphics[width=.45\linewidth,height=1.45in]{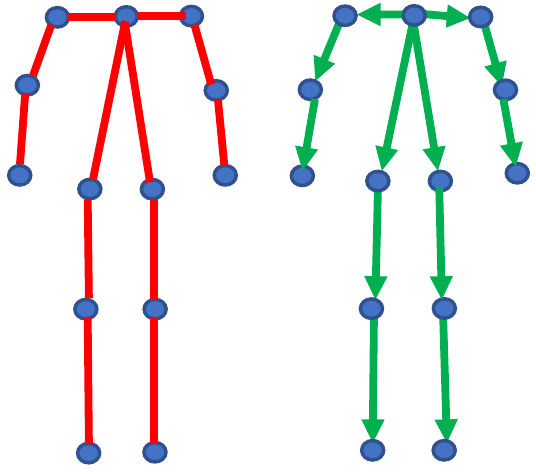}
\vspace{-.4cm}
\caption[model and features]{Left: Skeleton model, Right: features. Each green arrow shows one 3-dimensional vector in the feature vector.}
\label{fig:SkeletonModelAndFeatures}
\vspace{-.5cm}
\end{figure}
%\vspace{-.35cm}

Existing state-of-the-art model-based methods avoid the challenge of erroneous features by adopting high-quality skeleton data provided by Kinect or Mocap. In contrast, the data collected by a flash lidar camera is noisy and has low resolution that negatively affects the skeleton extraction performance. Faulty skeleton models result in features that are plagued with missing and erroneous measurements that in turn present a major challenge for a successful gait identification. This work takes on the challenging task of gait identification using flash lidar data. Our main contributions can be described as follows. First, we present the first model-based approach for gait recognition using flash lidar data. The only existing lidar-based person identification methods are model-free, and rely on silhouette extraction from a point cloud \cite{galai2015feature,benedek20143d}. Second, we propose a rule-based filtering mechanism to correct for erroneous skeleton joint coordinates by modeling each joint location as a time sequence and adopting robust statistics measures of the nearest neighbors. Third, experiments are performed using the flash lidar data to evaluate the performance of the proposed methodology.
\vspace{-.3cm}
\section{Proposed methodology}
\label{sec:methodology}
\vspace{-.3cm}
Figure \ref{fig:flowChart} demonstrates the workflow of the proposed methodology. For a video sequence like $V$ with $f$ frames, the input into the proposed gait recognition system are the intensity $I=[I_{1}, I_{2},...,I_{f}]$ and range data $R=[R_{1},R_{2},...,R_{f}]$, recorded by a single flash lidar camera, where images are preprocessed to reduce the noise. Figure \ref{fig:sampleFramesOfLidar} shows sample intensity and range data in the first and second row. OpenPose, a state-of-the-art real-time pose detector \cite{cao2016realtime}, is leveraged to extract a skeleton model from the intensity information of lidar. The employed skeleton model is illustrated in the left side of Figure \ref{fig:SkeletonModelAndFeatures}. For each frame, we present the 2-dimensional coordinates of skeleton joints in the vectorized form
\vspace{-.3cm}
\begin{equation}
\vspace{-.3cm}
\label{eq:OpenPoseOutput}
J_{i}=[x_{k},y_{k}]^{M_{j}}_{k=1}\in \Re^{2N} 
\end{equation}
where $M_{j}$ is the number of joints and $(x_{k},y_{k})$ are the coordinates of the $kth$ joint in the image frame of reference. The sample frames with the correct detected skeleton models can be seen in the third row of Figure \ref{fig:sampleFramesOfLidar}. Next, the range data is used to project the 2D locations of joints, provided by OpenPose, into a real-world coordinate system:
\vspace{-.3cm}
\begin{equation}
\vspace{-.3cm}
\label{eq:realWorld}
L^{i}_{j}  =  \frac{2}{N_{pixels}}\times\tan(\frac{\theta_{aov}}{2})\times Lp^{i}_{j} \times D^{i}_{camera}
\end{equation}
where $L^{i}_{j}$ is the real-world location of joint $i$ in the direction $j$ and $ Lp^{i}_{j}$ is the corresponding location in the image coordinate system. $N_{pixels}$ is the number of pixels in the $j$ direction, $\theta_{aov}$ is the angle of view, and $D^{i}_{camera}$ is the range value of joint $i$. Several factors in the data negatively affect the quality of features that are computed from the resulting joints. As the subjects loom closer to the camera, range data are affected by noise. The intensity data lack color, and there is similarity between human clothing, skin and the background. The last row in Figure \ref{fig:sampleFramesOfLidar} shows a few examples of the detected faulty skeletons, which are the result of noisy nature of lidar data and erroneous joint localization of OpenPose. Features that are computed from such faulty skeletons contain missing and erroneous values that presents a big challenge for a successful gait recognition. To resolve this problem, we present a filtering mechanism to correct joint location values and extract features after joint correction. Furthermore, to incorporate the dynamic of the motion, we perform feature concatenation with a new criterion. 
\begin{figure}[!ht]
\centering
\includegraphics[width=.95\linewidth]{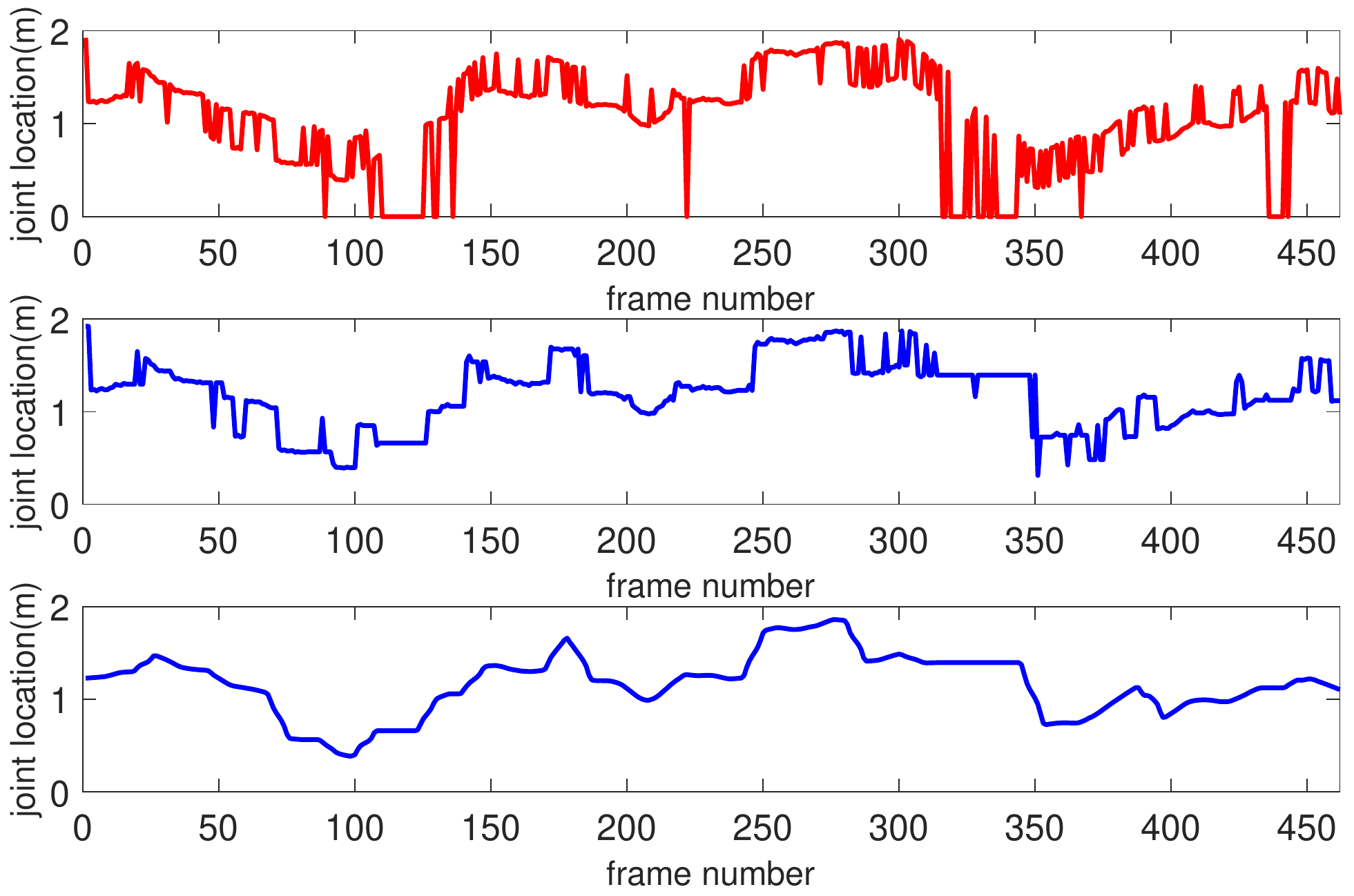}
\vspace{-.3cm}
\caption[joint correction]{Sample joint location time sequence in one direction before (top), after joint correction (middle), and after smoothing (bottom).}
\label{fig:SampleJointLocationB4AfterJointCorrection}
\vspace{-.8cm}
\end{figure}
%\vspace{-.2cm}
\section{Joint location correction}
\label{sec:JointCorrection}
\vspace{-.3cm}
The top row of Figure \ref{fig:SampleJointLocationB4AfterJointCorrection} shows a typical joint location time sequence in one direction. The significant number of missing values and sudden jumps in the joint location sequence results into features that are plagued with erroneous and outlier measurements. The outcome will be gait identification with lower accuracy. To resolve this problem, we propose a \textbf{rule-based short-memory median filter} to correct joint location time sequence. $L_{x}$, the joint location time sequence in the x direction, extended over $F_{n}$ frames, is defined as follows
\vspace{-.2cm}
\begin{equation}
\label{eq:JointLocation}
L_{x}=[L_{x}(t)]^{F_{n}}_{t=1} ~~~~~~L_{x}(t)\in \Re
\vspace{-.2cm}
\end{equation}
where $L_{y}$ and $L_{z}$ can be defined in the same way. We correct $L_{x}$ at time $t$, if there is a missing value ($L_{x}(t)=0$), or a sudden jump. We define the sudden jump using the relative distance of the current and previous values of the joint location
\vspace{-.25cm}
\begin{equation}
\label{eq:Threshold}
\frac{\left |L_{x}(t)-L_{x}(t-1)\right|}{L_{x}(t-1)}>median(\frac {\partial L_{x}}{\partial t})~~~~,\frac {\partial L_{x}}{\partial t} \neq 0
\vspace{-.25cm}
\end{equation}
Each instant corresponds with one frame, and $(\frac {\partial L_{x}}{\partial t})$ is the derivative of joint location in the \textit{x} direction with respect to time. If any of the above conditions are satisfied, the corrected joint location value at frame $t$ will be calculated according to the following equation:
\vspace{-.25cm}
\begin{equation}
\label{eq:CorrectedJoints}
\begin{cases}
       \text{$L_{x}(t)=median(W(t))$}\\
       \text{$W(t)=[L_{x}(i)\mid L_{x}(i) \neq 0]^{t-1}_{i=t-F}$}\\
       \text{$\mathbf{card}(W(t))=P_{nz},~~~~ F\geq P_{nz} $}\\
    \end{cases}     
\vspace{-.25cm}
\end{equation}
where $P_{nz}$ is the number of previous closest non-zero neighbors of joint location values at time $t$. $L_{x}(i)$ is the joint location value at previous closest non-zero neighbors of frame $t$. We define $W(t)$ as the array of closest nonzero neighbors at time $t$. $W(t)$ is updated continuously, containing at each instant the previous nonzero values of joint location sequence that are closest in time to the current instant. This includes the corrected joint location values.

The length of $W(t)$ is selected to follow the local pattern of $L_{x}(t)$. We used $\mathbf{card}(W(t))=3$ as the smallest possible number of previous nearest neighbors. $\mathbf{card}(W(t))=1$ and $\mathbf{card}(W(t))=2$ are not proper choices for this problem. Choosing $\mathbf{card}(W(t))=1$ corresponds with the previous nearest non-zero neighbor. However, if the nearest non-zero neighbor is noisy, the error will propagate as a result of correction. On the other hand, considering $\mathbf{card}(W(t))=2$ for the length of $W$, transforms median into mean. Median is selected as a robust statistic that is less affected by outliers. In this problem, by choosing median over mean, the effect of erroneous joint location values can be ameliorated. Furthermore, median filtering is effective in removing the impulse-like signal features, which occur in this application due to be the pattern of missing joints and jumps in the joint location sequence. The middle row of Figure \ref{fig:SampleJointLocationB4AfterJointCorrection} shows a sample joint location sequence along one direction after applying joint correction.

\begin{figure}[!ht]
\centering
\includegraphics[width=.95\linewidth]{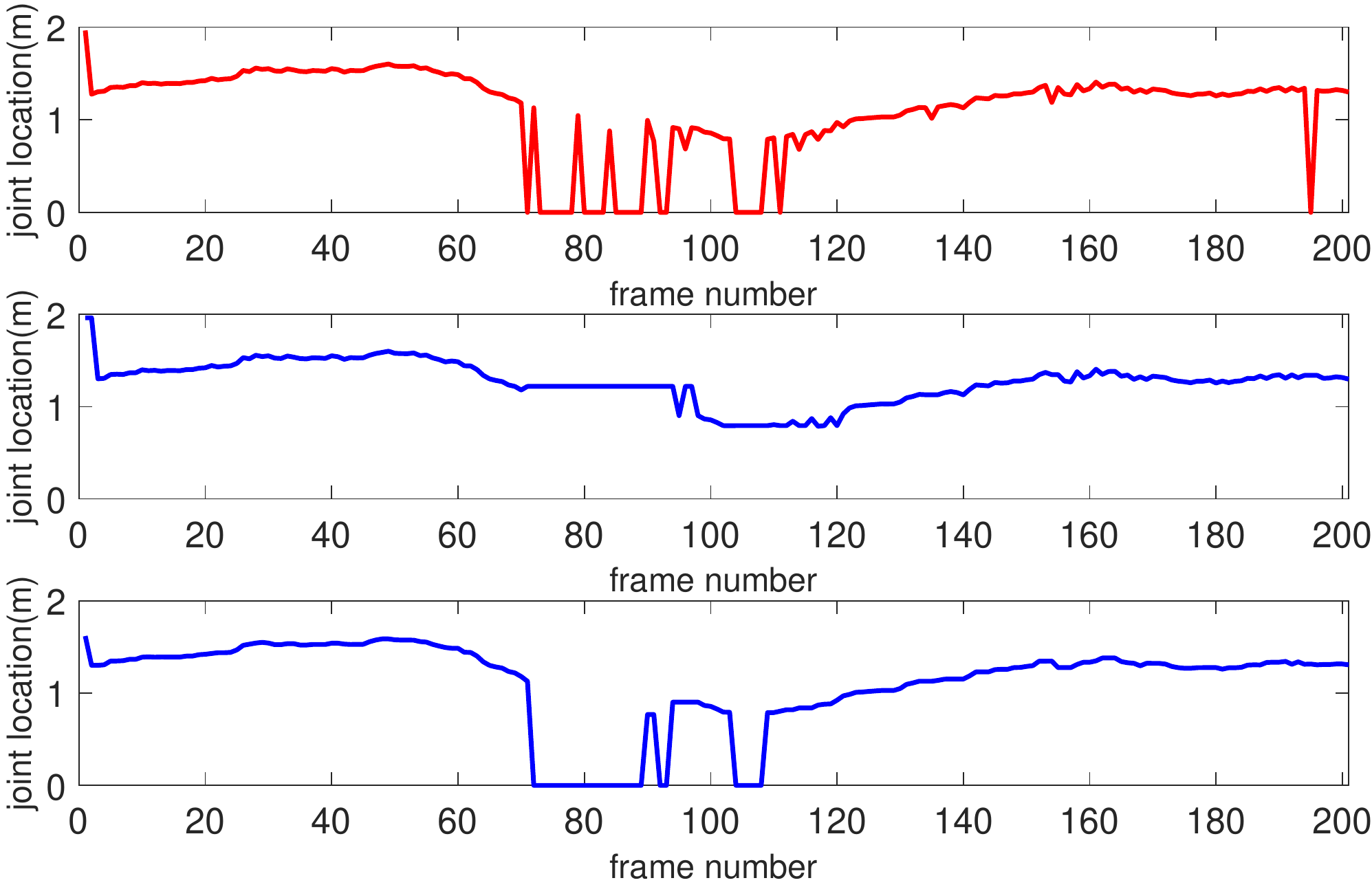}
\vspace{-.4cm}
\caption[Our Moving Median Vs Moving Median Filter]{From top to bottom: original joint location in one direction, after joint correction with the proposed rule-based median, after applying moving median filter of the same window length}
\label{fig:ComparingTheMovingWindowAndMovMedian}
\vspace{-.45cm}
\end{figure}

Figure \ref{fig:ComparingTheMovingWindowAndMovMedian} shows a comparison between the proposed rule-based median filter with the moving median. Our rule-based filter is in particular successful in correcting missing values when they occur over consecutive frames. This is mainly the result of the way each of the above two filters work. In particular, the moving median uses the neighborhood information irrespective of their values. In contrast, our rule-based median filter uses the previous neighbors' values only if they are nonzero and there is no sudden jump between consecutive values. However, as can be seen in Figure \ref{fig:SampleJointLocationB4AfterJointCorrection} and \ref{fig:ComparingTheMovingWindowAndMovMedian}, this can also cause the flattening of the signal in some regions.
\begin{table}[!t]
\caption[feature Vector]{List of three-dimensional vectors in the feature vector (L refers to the left joints and R refers to the right joints)}
\vspace{-.35cm}
\label{table:featureVector}
\resizebox{\columnwidth}{!}{%
\begin{tabular}{|c|c|c|c|c|}
\hline
\rule{0pt}{12pt} Neck to R Shoulder & Neck to L Shoulder & Neck to R Hip & Neck to L Hip\\
\hline
\rule{0pt}{12pt} R Shoulder to R Elbow & L Shoulder to L Elbow & R Hip to R Knee & L Hip to L Knee\\
\hline
\rule{0pt}{12pt} R Elbow to R Wrist & L Elbow to L Wrist & R Knee to R Ankle & L Knee to L Ankle\\
\hline
\end{tabular}%
}
\centering
\vspace{-.5cm}
\end{table}

Finally, in order to alleviate the effects of signal flattening and lower amplitude impulses, both the result of joint correction in regions with consecutive missing values or sudden jumps, we employ RLowess (locally weighted scattered plot smoothing) \cite{cleveland1979robust}, that locally fits first order polynomial using weighted linear regression, where regression weights are estimated through a robust procedure. The robustness of the employed method is essential due to the existence of low-amplitude impulses that act as outliers. In Figure \ref{fig:SampleJointLocationB4AfterJointCorrection} the last row shows the smoothed joint location time sequence.
\vspace{-.5cm}
%\vspace{-.1cm}
\section{feature vectors}
\label{sec:Features}
\vspace{-.3cm}
Like \cite{kumar2012human}, our features is comprised by a set of 3-dimensional vectors measured between selected joints of skeleton model. Compared to features that describe the distance between joints \cite{yang2016relative}, or to features that only consider angles between selected joints \cite{ball2012unsupervised}, we can implicitly encode both the distance, and the angles of selected joints in the skeleton in different postures. Each three-dimensional vector $V_{ij}$ in our feature vector is defined according to equation \ref{eq:setOf3DFeature}:
\begin{equation}
\vspace{-.2cm}
\label{eq:setOf3DFeature}
V_{ij}  =  (X_{i} - X_{j} , Y_{i} - Y_{j} , Z_{i} - Z_{j}),~~ where~~ i \neq j 
\end{equation}
where $i$ and $j$ are the indices of selected joints. Unlike features in \cite{kumar2012human} that were defined with respect to a reference joint, our features are formulated between different joints. Table \ref{table:featureVector} describes the name of the joints that form each of the three-dimensional vectors. Figure \ref{fig:SkeletonModelAndFeatures}, right illustrates the described vectors.
\begin{table}[!ht]
\vspace{-.2cm} 
\caption[results]{Correct identification scores (accuracy and F-score) for the proposed features (**), the methods in \cite{sinha2013person} and \cite{yang2016relative}. Features are computed without joint correction.}
\vspace{-.3cm}
\label{table:ResultsComparisonOriginalData}
\resizebox{.72\columnwidth}{!}{
\begin{tabular}{|c|c|c|c|c|}
\hline
  & \cite{sinha2013person} & \cite{yang2016relative} & ** \\
\hline
Accuracy & 43.07\% & 45.67\% & \textbf{56.26}\% \\
\hline
F-Score & 42.41\% & 43.72\% & \textbf{57.24}\%\\
\hline
\end{tabular}
}
\vspace{-.5cm}
\end{table}

\begin{table}[!ht]
\caption[results]{Correct identification scores for $\mathcal{G}$lidar3DJ, $\mathcal{G}$lidar3DJ with feature concatenation (F-C), and the methods in \cite{sinha2013person} and \cite{yang2016relative}. Features are computed from corrected joints.}
\vspace{-.3cm}
\label{table:ResultsComparisonJointCorrection}
\resizebox{\columnwidth}{!}{
\begin{tabular}{|c|c|c|c|c|c|}
\hline
\rule{0pt}{9pt}  &\cite{sinha2013person}  & \cite{yang2016relative} & $\mathcal{G}$lidar3DJ &  $\mathcal{G}$lidar3DJ(F-C)  \\
\hline
\rule{0pt}{9pt} Accuracy & 61.20\% &  70.59\% & 81.24\% &\textbf{85.11}\% \\
\hline
\rule{0pt}{9pt} F-Score & 57.41\% & 65.15\%  & 80.30\% & \textbf{84.33}\% \\
\hline
\end{tabular}
}
\centering
\vspace{-.4cm}
\end{table}
\vspace{-.4cm}
\section{feature concatenation}
\label{sec:Concatenation}
\begin{figure}[!ht]
\centering
\includegraphics[width=.95\linewidth]{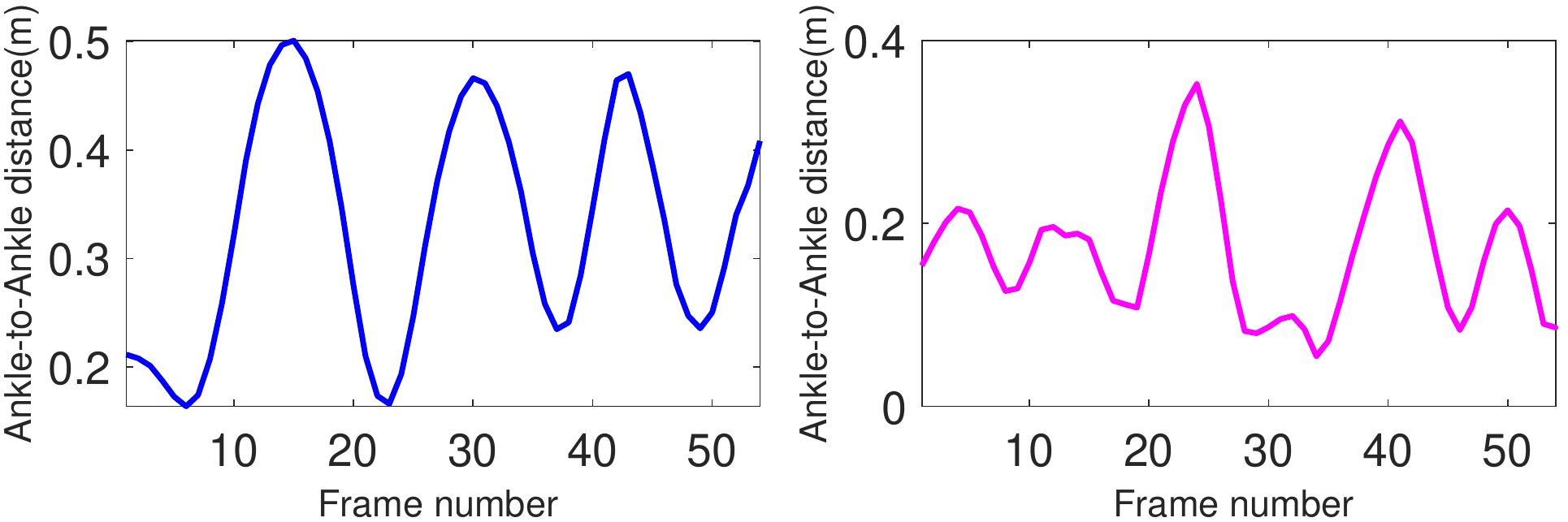}
\vspace{-.2cm}
\caption[Ankle-to-ankle]{Ankle-to-ankle distance for Kinect (left) vs. lidar(right).}
\label{fig:KinectVsLidar}
\vspace{-.4cm}
\end{figure}

\vspace{-.3cm}
When two individuals have similar body measurements in multiple postures, motion dynamics can play a crucial role in gait identification. A common practice to include motion dynamics in many model-based methods is to compute features like speed and step length using ankle-to-ankle distance sequence \cite{preis2012gait, koide2016identification}, or to calculate moments like variance, maximum and average of selected features in each gait cycle \cite{sinha2013person,yang2016relative,chi2018gait}. The gait cycle is estimated by looking at the distance between the two ankle joints time sequence, which is straightforward with clean joint position data. Figure \ref{fig:KinectVsLidar} shows a comparison between ankle-to-ankle distance sequences, one from Kinect, and one from joint-corrected samples of our lidar data set. As can be seen in this figure, while a clear cyclic pattern can be observed for the Kinect sample, it is difficult to determine a gait cycle in the lidar time sequence. To compensate for such shortcomings, instead of calculating feature moments, we concatenate feature vectors in the consecutive frames to encode the dynamics of the motion.

To find a proper window length for feature concatenation, we use the idea of gait cycle; however, the gait cycle can vary from subject to subject and even during a course of walking sequence. To resolve this issue, we first remove the small and large values of gait cycle from each of the ankle-to-ankle distance curve. These small and large values occur in the beginning and end of the motion, as well as when the subjects change their motion direction. Once such outlier gait cycles are removed, majority voting is performed on the remaining gait cycles. The gait cycle that appears the most is selected as the length of window for feature concatenation.
\vspace{-.3cm}
\vspace{-.4cm}
\section{Results}
\label{sec:majhead}
\vspace{-.3cm}
The data set recorded by a single flash lidar camera includes 34 sequences of walking from 10 subjects in which the camera is fixed during all the actions. The walking action is performed in three different ways, capturing multiple views of the subjects: walking toward and away from the camera, walking on a diamond shape, and walking on a diamond shape while holding a yard stick in one hand. We used 70\% of the sequences for training and the rest for testing, where the classifier is tested on a type of walking that it was not trained on. K-nearest neighbors of $N = 7$ and Manhattan distance is adopted as our classifier. The performance of the proposed approach is compared with the works in \cite{sinha2013person} and \cite{yang2016relative}, which are among state-of-the-art model-based methods. Table \ref{table:ResultsComparisonOriginalData} shows the correct identification scores with the proposed features, the features in \cite{sinha2013person} and \cite{yang2016relative} without joint correction. Results in Table \ref{table:ResultsComparisonJointCorrection} report identification scores with the proposed joint correction. It also shows the scores after applying feature concatenation on $\mathcal{G}$lidar3DJ. By comparing the results, it is clear that joint correction can drastically improve gait identification accuracy in all of the cases. The results also demonstrate the advantage of feature concatenation over the model-based approaches that rely on a combination of static anthropometric-based attributes, and statistical moments that describe dynamic features \cite{sinha2013person,yang2016relative}.
\vspace{-.3cm}
\section{Conclusion}
\label{sec:conclusion}
\vspace{-.3cm}
In this work, we introduce $\mathcal{G}$lidar3DJ, the \textit{first} model-based approach for gait recognition using flash lidar data. Our model-based approach is scale and view invariant, which is essential for real-world application. We address a major limitation of the current state-of-the-art model-based methods that require high quality Mocap or Kinect data. The noisy, low resolution flash lidar data used in this study present challenges to the performance of state-of-the-art skeleton detectors, degrading joint localization and gait identification accuracy. A rule-based short-memory median filter is presented that improves the quality of features and gait recognition accuracy. The proposed median filter has the potential application for predicting missing values in time series. Furthermore, we introduce a new feature concatenation criterion to incorporate the dynamic of motion and improve gait recognition methodology. Experimental results support the effectiveness of $\mathcal{G}$lidar3DJ for gait recognition despite noisy data with faulty missing features.
\bibliographystyle{IEEEbib}
\small
%\vspace{-1.12cm}
\bibliography{Template}

\end{document}